\documentclass[showpacs,showkeys,preprintnumbers, amsmath,amssymb,preprint]{revtex4-1}
\usepackage{amssymb}
\usepackage{amsfonts}
\usepackage{amsmath}
\usepackage{graphicx}

\usepackage{graphicx}

\begin{document}

\title{Quasinormal modes of non-Abelian hyperscaling violating Lifshitz black holes}
\author{Ram\'{o}n B\'{e}car}
\email{rbecar@uct.cl}
\affiliation{Departamento de Ciencias Matem\'{a}ticas y F\'{\i}sicas, Universidad Cat\'{o}%
lica de Temuco, Montt 56, Casilla 15-D, Temuco, Chile}
\author{P. A. Gonz\'{a}lez}
\email{pablo.gonzalez@udp.cl}
\affiliation{Facultad de Ingenier\'{\i}a, Universidad Diego Portales, Avenida Ej\'{e}%
rcito Libertador 441, Casilla 298-V, Santiago, Chile.}
\author{Yerko V\'{a}squez}
\email{yvasquez@userena.cl}
\affiliation{Departamento de F\'{\i}sica y Astronom\'{\i}a, Facultad de Ciencias, Universidad de La Serena,\\
Avenida Cisternas 1200, La Serena, Chile.}
\date{\today}

\begin{abstract}

We study the quasinormal modes of scalar field perturbations in the background of non-Abelian hyperscaling violating Lifshitz black holes. We find that  the quasinormal frequencies have no real part so there is no oscillatory behavior in the perturbations, only exponential decay, that is, the system is always overdamped, which guarantees the mode stability of non-Abelian hyperscaling violating Lifshitz black holes. We determine analytically the quasinormal modes for massless scalar fields for a dynamical exponent $z=2$ and hyperscaling violating exponent $\tilde{\theta}>-2$. Also, we obtain numerically the quasinormal frequencies for different values of the dynamical exponent and the hyperscaling violating exponent by using the improved asymptotic iteration method.

\end{abstract}

\maketitle


\tableofcontents


\newpage
\section{Introduction}

The gauge/gravity duality contains interesting gravity theories. One of them is the known anti-de Sitter (AdS) gravity theory, which can be dual to conformally invariant field theories. Another theory, known as Lifshitz gravity, can be dual to scale-invariant field theories, being non-conformally invariant which present an anisotropic scale invariance $t\rightarrow \lambda ^{z}t$, $x_i\rightarrow \lambda x_i$, 
where the dynamical exponent $z$ is the relative scale dimension of time $t$ and space $x_{i}$. Spacetimes that exhibit these symmetries along with a scaling $r\rightarrow \lambda^{-1} r$ for the radial coordinate
are known as Lifshitz spacetimes and are described by the following metrics \cite{Kachru:2008yh} 
\begin{equation}
ds^2=- \frac{r^{2z}}{\ell^{2z}}dt^2+\frac{\ell^2}{r^2}dr^2+\frac{r^2}{\ell^2} d\vec{x}%
^2~,  \label{lif1}
\end{equation}
where $\vec{x}$ represents a $D-2$ dimensional spatial vector, $D$ is the
spacetime dimension and $\ell$ denotes the length scale in these geometries.
Also, by introducing both an Abelian gauge field and a scalar dilaton, spacetimes emerge which, in addition to having an anisotropic scaling exponent $z$ as the Lifshitz metric, have an overall hyperscaling violating factor with hyperscaling exponent $\theta$ that is not scale invariant; thus, this line element is conformally related to the Lifshitz metric and transforms as $ds\rightarrow \lambda ^{\frac{\theta}{D-2}}ds$ under the Lifshitz scaling. The metric can be represented as:
\begin{equation}
ds^2= r^{-\frac{2 \theta}{D-2}} \left(- \frac{r^{2z}}{\ell^{2z}}dt^2+\frac{\ell^2}{r^2}dr^2+\frac{r^2}{\ell^2} d\vec{x}%
^2 \right)~.
\end{equation}
This spacetime is important in the study of dual field theories with hyperscaling violation  \cite{Dong:2012se, Narayan:2012hk, Perlmutter:2012he, Ammon:2012je, Bhattacharya:2012zu, Dey:2012fi, Alishahiha:2012qu, Gath:2012pg, Bueno:2012vx, Iizuka:2012pn, Fan:2013tpa, Fan:2013zqa, Fan:2013tga, Bhattacharya:2014dea}, and it might be a gravitational representation of  a theory with a Fermi surface  in terms of its leading large $N$ thermodynamic behavior \cite {Ogawa:2011bz, Huijse}. On the other hand, Lifshitz black holes with hyperscaling violation have been found in \cite{Dehghani:2015gza, Ganjali:2015cba}. Moreover, Einstein gravity coupled to a cosmological constant and multiple $SU(2)$ Yang-Mills fields
admits colored Lifshitz solutions as well as
hyperscaling violating Lifshitz black holes when we introduce a dilaton to the system and a Maxwell field \cite{Feng:2015yja}.\\

The study of quasinormal modes (QNMs) and their quasinormal frequencies (QNFs) \cite{Regge:1957td, Zerilli:1971wd,
Zerilli:1970se, Kokkotas:1999bd, Nollert:1999ji, Konoplya:2011qq} provides
information about the stability of matter fields that
evolve perturbatively in the exterior region of a black hole without backreacting on the
metric. They are independent of the initial conditions and depend only on the parameters of the black hole (mass, charge and angular momentum) and the fundamental constants (Newton constant and cosmological constant) that describe
a black hole, just like the parameters that define the test field.
On the other hand, QNFs determine how fast a thermal state in the
boundary theory will reach thermal equilibrium according to the AdS/CFT
correspondence \cite{Maldacena:1997re}, where the relaxation time of a
thermal state is proportional to the inverse
of the imaginary part of the QNFs of the dual gravity background, which was
established due to the QNFs of the black hole being related to the poles of
the retarded correlation function of the corresponding perturbations of the
dual conformal field theory \cite{Birmingham:2001pj}. The QNFs have been calculated by means of numerical and analytical techniques; some remarkable numerical methods are: the Mashhoon method, the Chandrasekhar-Detweiler, the WKB method, the Frobenius method, the method of continued fractions, the Nollert, the asymptotic iteration method (AIM) and the improved AIM, among others. In this work, we will perform numerical studies
using the improved AIM \cite{Cho:2009cj}, which is an improved version of the method proposed in references \cite{Ciftci, Ciftci:2005xn} and which has been applied successfully in the context of QNMs for different black hole geometries (see for instance \cite{Cho:2009cj, Cho:2011sf, Catalan:2013eza, Catalan:2014ama, Zhang:2015jda, Barakat:2006ki, Sybesma:2015oha, Gonzalez:2015gla}). QNMs of Lifshitz black holes under scalar field perturbations have been studied in \citep{CuadrosMelgar:2011up, Gonzalez:2012de, Gonzalez:2012xc, Myung:2012cb, Becar:2012bj,Giacomini:2012hg, Lepe:2012zf, Catalan:2014ama, Catalan:2014una}, and generally the scalar modes of Lifshitz black holes are stable, with the imaginary part being negative and the modes decaying in time. Moreover, it  was established that for a black brane solution of an Einstein-Maxwell-Dilaton action \cite{Taylor:2008tg, Tarrio:2011de}, for $d> z+1$, at zero momenta, the modes are non-overdamped, whereas for $d \leq z + 1$ the system is always overdamped, $d=D-1$ being the boundary dimension \cite{Sybesma:2015oha}; that is, the QNFs have no real part so there is no oscillatory behavior in the perturbations, only exponential decay. A similar behavior was observed in a topological nonlinearly charged Lifshitz black hole as a solution for the Einstein-dilaton gravity in the presence of a power-law and two linear Maxwell electromagnetic fields, where the modes are overdamped depending heavily on the dynamical exponent and the angular momentum of the scalar field for a spherical transverse section \cite{Becar:2015kpa}. Also, in the context of QNMs of black branes with hyperscaling violation, it was established that the scalar and electromagnetic modes of a black brane are both stable with zero spatial momentum \cite{BAI:2013koa}. Moreover, Einstein-dilaton gravity theory in the presence of a linear and a nonlinear electromagnetic field has a nonlinear charged Lifshitz black brane with hyperscaling violation as a solution  \cite{Dehghani:2015gza}. In this case, it was shown that the modes are overdamped for $D\leq z+2+\theta$, whereas for $D>z+2+\theta$  the modes are non-overdamped in all the cases analyzed \cite{Gonzalez:2015gla}. Non-relativistic fermion Green's functions in 4-dimensional Lifshitz spacetime with $z=2$ were studied in \cite{Alishahiha:2012nm} by considering fermions in this background and a non-relativistic (mixed) boundary condition, and it was shown, among other things, that the Green's functions have a flat band. Also, the Dirac QNMs of a 4-dimensional Lifshitz black hole in \cite{Catalan:2013eza} and the electromagnetic QNMs in \cite{Lopez-Ortega:2014oha} were studied. In the context of black hole
thermodynamics, QNMs make it possible to study the quantum area spectrum of the black hole
horizon \cite{CuadrosMelgar:2011up} as well as the mass and
the entropy spectrum. Additionally, the scalar greybody factors for an asymptotically Lifshitz black hole were studied in \cite{Gonzalez:2012xc, Lepe:2012zf}, and particle motion on these geometries in \cite{Olivares:2013zta, Olivares:2013uha, Villanueva:2013gra}.\\

In this work, we study scalar perturbations of non-Abelian charged Lifshitz black holes with hyperscaling violation.
The matter is parameterized by scalar fields minimally coupled to gravity. Then, we obtain the quasinormal frequencies for scalar fields analytically and numerically. We focus our study on the influence of the dynamical  exponent and the hyperscaling exponent on the frequencies. As we will show, the QNFs have a negative imaginary part and are always overdamped, satisfying  $D \leq z+2-(D-2) \tilde{\theta} /2$, where $ \tilde{\theta}=-\frac{2\theta}{D-2} $ corresponds to the hyperscaling exponent for the geometry under study. Therefore, our results are consistent with the aforementioned geometries.\\

The manuscript is organized as follows: In Sec. \ref{Background} we give a brief review
 of a non-Abelian hyperscaling violating Lifshitz black hole.
In Sec. \ref{QNM}
we calculate the QNFs of scalar perturbations analytically and numerically by using the improved AIM. Finally, our conclusions are in Sec. \ref{conclusion}.

\section{Non-Abelian hyperscaling violating Lifshitz black holes}
\label{Background}
The non-Abelian charged Lifshitz black hole that we consider is a solution of the Einstein gravity coupled to a cosmological constant, $N SU(2)$ Yang-Mills fields $A_I^a$ $(a=1, 2, 3$ and $I= 1, 2, ... , N)$, Maxwell field $\mathcal{A}= \bar\varphi dt$ and a dilaton field $\phi$  \cite{Feng:2015yja}. The Lagrangian is
\begin{equation}\label{action}
\mathcal{L}_D=\sqrt{-g}\left(R-V(\phi)-\frac{1}{2}(\partial \phi)^2-\sum_{I=1}^N \frac{1}{4g_I^2}e^{\lambda \phi}F_I^2-\frac{1}{4}e^{\lambda \phi}\mathcal{F}^2 \right)~,
\end{equation}
where $\mathcal{F}^2=F^a_{\mu\nu}F^{a\mu\nu}$ and $g_I$ is the coupling constant of the Yang-Mills term in the action. The Yang-Mills and Maxwell field strengths are defined, respectively, as
\begin{equation}
F^a_{\mu\nu}=\partial_\mu A^a_\nu-\partial_\nu A^a_\mu+\epsilon^{abc}A^b_\mu A^c_\nu~,\mathcal{F}=d\mathcal{A}~.
\end{equation}
The following metric is a solution of the theory defined by the Lagrangian (\ref{action})
\begin{equation}\label{metric}
ds^2=r^{\tilde{\theta}} \left(-r^{2z}f(r)dt^2+\frac{dr^2}{r^2 f(r)}+r^2 \sum _{i=1}^{D-2} dx_{i}^2 \right)~,
\end{equation}
where
\begin{equation}
f(r)=1-\frac{q^{2}}{2(z-1)r^{2(z-1)}}~,
\end{equation}
\begin{equation}\label{hyp}
\tilde{\theta}=\frac{2}{D-2}[z-(D-1)]~,
\end{equation}
\begin{equation}
\lambda=\sqrt{\frac{2z-2(D-1)}{(D-1)(D-2)(z-1)}}~,\Lambda=1-z-D(z-1)^2~, V(\phi)=\Lambda e^{-\lambda\phi}~,
\end{equation}
and the fields are given by
\begin{equation}
\psi=\sqrt{z-1}r~, \bar\varphi= \bar\varphi_0+qr~,\phi=\sqrt{\frac{2(D-1)}{D-2}(z-1)[z-(D-1)]}~\log~r~,
\end{equation}
where $\psi(r)$ is a function related to the potentials of the Yang-Mills fields \cite{Feng:2015yja}, $q$ is an integration constant proportional to the electric charge of the black hole and $\bar{\varphi}_{0}$ is a constant. The metric for $z>1$ represents a non-Abelian charged Lifshitz black hole solution with hyperscaling violating factor $r^{\tilde{\theta}}$. The hyperscaling exponent $\tilde{\theta}$ is constrained in Eq. \eqref{hyp}, and this constraint implies that $\tilde{\theta}>-2$ for $z>1$. Furthermore, note that when $z<D-1$, the dilaton field and the coupling constant $\lambda$ become purely imaginary. However, the reality of the Lagrangian (\ref{action}) can be restored by letting $\phi \rightarrow i\phi$. This implies that the dilaton $\phi$ for $z<D-1$ has the wrong kinetic sign and is therefore ghost-like. Also, when $\lambda=0$, corresponding to $z=D-1$, we have $\tilde{\theta}=0$, and the dilaton decouples from
the theory \cite{Fan:2015aia}.

\section{Scalar perturbations }
\label{QNM}
\begin{figure}[h]
\begin{center}
\includegraphics[width=0.9\textwidth]{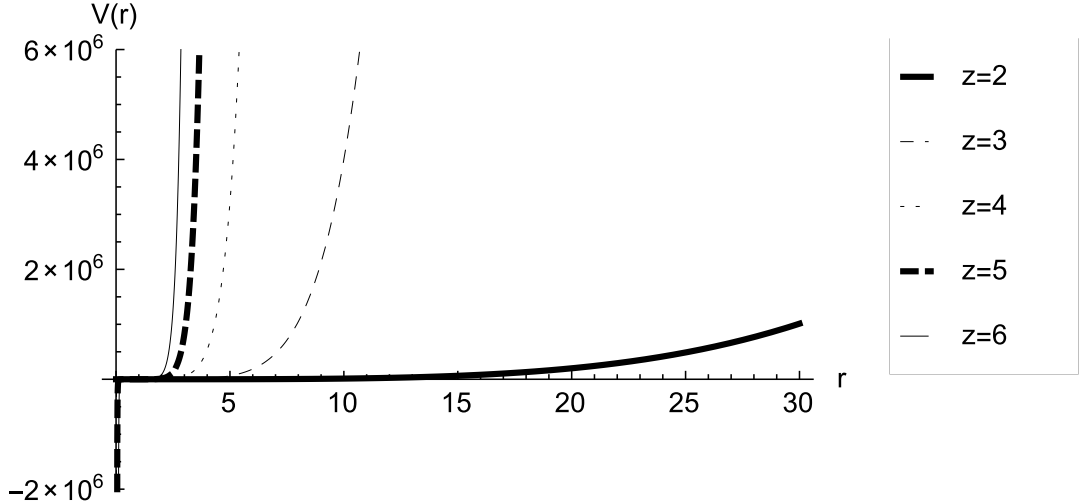}
\end{center}
\caption{The behavior of $V(r)$ with $m=0$, $q=1$, $\kappa=0$, and $z =2,3,4,5,6$.} \label{Potential1}
\end{figure}

The QNMs of scalar perturbations in the background of non-Abelian hyperscaling violating Lifshitz black holes are given by the scalar field solution of the Klein-Gordon equation
\begin{equation}
\frac{1}{\sqrt{-g}}\partial _{\mu }\left( \sqrt{-g}g^{\mu \nu }\partial
_{\nu } \varphi \right) =m^{2}\varphi \,,  \label{KGNM}
\end{equation}%
with suitable boundary conditions for a black hole geometry. In the above expression $m$ is the mass of the scalar field $\varphi $. Now, by means of the following ansatz
\begin{equation}
\varphi =e^{-i\omega t}e^{i\vec{\kappa} \cdot \vec{x}}R(r)\,,
\end{equation}%
where $\vec{x}$ is a $(D-2)$-dimensional spatial vector and $-\kappa^2$ is the eigenvalue of the Laplacian in the flat base submanifold, the Klein-Gordon equation reduces to
\begin{equation}
 \frac{1}{r^{\eta}}\frac{d}{dr}\left(r^{2+\eta-\tilde{\theta}}f(r)\frac{dR}{dr}\right)+\left(\frac{\omega^{2}}{r^{\tilde{\theta}+2z}f(r)}-\frac{\kappa ^2}{r^{2+\tilde{\theta}}}-m^{2}\right) R(r)=0\,, \label{radial}
\end{equation}%
where we have defined $\eta=\frac{\tilde{\theta} D}{2}+z+D-3$. Now, defining $R(r)$ as
 \begin{equation}
 R(r)=\frac{F(r)}{r^{s}}\,,
 \end{equation}
where $s=\frac{(D-2)(1+\tilde{\theta}/2)}{2}$, and by using the tortoise coordinate $r^*$ given by
 \begin{equation}\label{ast}
 dr^*=\frac{dr}{r^{z+1}f(r)}\,,
 \end{equation}
 the Klein-Gordon equation can be written as a one-dimensional Schr\"{o}dinger equation
 \begin{equation}\label{ggg}
 \frac{d^{2}F(r^*)}{dr^{*2}}-V(r)F(r^*)=-\omega^{2}F(r^*)\,,
 \end{equation}
 with an effective potential $V(r)$, which is parametrically thought of as $V(r^*)$, given by
  \begin{equation}\label{pot}
 V(r)=r^{2z-2}f(r) \left(s r^{3}\frac{df}{dr}+s(s+z)r^{2}f(r)+\kappa^2+m^2 r^{2+\tilde{\theta}} \right)~.
 \end{equation}
Substituting the function $f(r)$ and the parameters $s$ and $\tilde{\theta}$ in the above expression we get
\begin{equation}
V(r)=\frac{1}{16(z-1)}(2r^{2z}(z-1)-q^2r^2) \left(2-8z+6z^2+q^2 r^{2-2z} (z-3)+8\frac{\kappa^2}{r^2}+8m^2r^{\frac{2-2D+2z}{D-2}} \right).
\end{equation}
From this expression we note that for massless scalar fields, the effective potential and thus Eq. (\ref{ggg}) turn out to be independent of the spacetime dimension. This implies, as we shall see later, that the QNFs ($\omega$) will also not depend on the spacetime dimension for massless scalar fields. The effective potential diverges at spatial infinity, see Fig. \ref{Potential1}, where we have considered massless scalar fields with $q=1$, $\kappa=0$, and $z =2, 3, 4, 5, 6$ which corresponds when $D=4$ to $\tilde{\theta}=-1, 0, 1, 2, 3$, respectively.

In the vicinity of the event horizon $r_h$, $f(r) \approx f'(r_h)(r-r_h)$ and from \eqref{ast} the tortoise coordinate is given by

\begin{equation}
r^* \approx \frac{1}{r_h^{z+1}f'(r_h)}\ln (r-r_h),
\end{equation}
so, $r^* \rightarrow -\infty$ when $r \rightarrow r_h$. On the other hand, at the spatial infinity $r \rightarrow \infty$, $f(r) \rightarrow 1$, and the tortoise coordinate is given by $r^* \approx -\frac{1}{z r^z} \rightarrow 0$.

\subsubsection{Exact quasinormal modes for $z=2$ black holes}
In this section we are interested in studying the scalar quasinormal ringing in the background \eqref{metric} in the particular case of $z=2$, because in this case the problem turns out to be exactly solvable in terms of hypergeometric function.
The Klein-Gordon equation (\ref{radial}) can be written as
\begin{eqnarray}
\notag && x\partial^2_xR(x)+\left(\frac{3}{2}-z-\frac{(z-1)x^{z-1}}{(1-x^{z-1})}\right)\partial_xR(x)+\\
&&\left(\frac{\omega^2 x^{z-1}}{4r_h^{2z}(1-x^{z-1})^2}-\frac{\kappa^2}{4(1-x^{z-1})r_h^2}-\frac{m^2r_h^{\tilde{\theta}}}{4x^{1+\tilde{\theta}/2}(1-x^{z-1})}\right)R(x)=0~,
\end{eqnarray}
where we have used $x=(r_h/r)^2$. Now, by considering
massless scalar field perturbations in the background of a non-Abelian charged Lifshitz black hole with hyperscaling violation and with dynamical exponent $z=2$, the above equation can be rewritten as
\begin{equation}
 x\partial^2_xR(x)-\left(\frac{1}{2}+\frac{x}{1-x}\right)\partial_xR(x)+\left(\frac{\omega^2 x}{4r_h^{4}(1-x)^2}-\frac{\kappa^2}{4(1-x)r_h^2}\right)R(x)=0~,
\end{equation}
which under the change of variable $x=1-y$ yields
\begin{equation}
 y(1-y)\partial^2_yR(y)+\left(1-\frac{y}{2}\right)\partial_yR(y)+\left(\frac{\omega^2 (1-y)}{4r_h^{4}y}-\frac{\kappa^2}{4r_h^2}\right)R(y)=0~,
\end{equation}
and if in addition we define $R(y)=y^{\alpha}F(y)$, the above equation leads to the hypergeometric equation
\begin{equation}\label{hypergeometric}
 y(1-y)F''(y)+\left[c-(1+a+b)y\right]F'(y)-ab F(y)=0~,
\end{equation}
 where
\begin{equation}
\alpha=  \pm\frac{i\omega}{2r_h^2}~,
\end{equation}
and the constants are given by
\begin{equation}\label{a}
a_{1,2}= \alpha -\frac{1}{4} \mp \frac{\sqrt{ r_h^4-4 r_h^2 \kappa^2-4 \omega^2}}{4 r_h^2}~,
\end{equation}
\begin{equation}
b_{1,2}= \alpha -\frac{1}{4} \pm \frac{\sqrt{ r_h^4-4 r_h^2 \kappa^2-4 \omega^2}}{4 r_h^2}~,\end{equation}
\begin{equation}
c=1+2\alpha~.
\end{equation}
The general solution of the hypergeometric equation~(\ref{hypergeometric}) is
\begin{equation}
\label{HSolution}
F(y)=C_{1}{_2}F{_1}(a,b,c;y)+C_2y^{1-c}{_2}F{_1}(a-c+1,b-c+1,2-c;y)~,
\end{equation}
and it has three regular singular points at $y=0$, $y=1$, and
$y=\infty$. ${_2}F{_1}(a,b,c;y)$ is a hypergeometric function
and $C_{1}$ and $C_{2}$ are integration constants.
So, in the vicinity of the horizon, $y=0$, and using
the property $F(a,b,c,0)=1$, the function $R(y)$ behaves as
\begin{equation}\label{Rhorizon}
R(y)=C_1 e^{\alpha \ln y}+C_2 e^{-\alpha \ln y},
\end{equation}
and the scalar field $\varphi$, for $\alpha=\alpha_-$ can be written as follows:
\begin{equation}
\varphi\sim C_1 e^{-i\omega (t+ \ln y /(2r_h^2))}+C_2
e^{-i\omega (t-\ln y/(2r_h^2))}~,
\end{equation}
in which the first term represents an ingoing wave and the second an outgoing wave in the black hole. So, by imposing that
only ingoing waves exist on the event horizon, this fixes  $C_2=0$. The radial
solution then becomes
\begin{equation}\label{horizonsolution}
R(y)=C_1 e^{\alpha \ln y}{_2}F{_1}(a,b,c;y)=C_1e^{-\frac{i\omega}{2r_h^2} \ln y}{_2}F{_1}(a,b,c;y)~.
\end{equation}
To implement boundary conditions at infinity ($y=1$), we
apply Kummer's formula
for the hypergeometric function \cite{M. Abramowitz},
\begin{equation}\label{relation}
{_2}F{_1}(a,b,c;y)=\frac{\Gamma(c)\Gamma(c-a-b)}{\Gamma(c-a)\Gamma(c-b)}F_1+(1-y)^{c-a-b}\frac{\Gamma(c)\Gamma(a+b-c)}{\Gamma(a)\Gamma(b)}F_2~,
\end{equation}
where
\begin{equation}
F_1={_2}F{_1}(a,b,a+b-c;1-y)~,
\end{equation}
\begin{equation}
F_2={_2}F{_1}(c-a,c-b,c-a-b+1;1-y)~.
\end{equation}
With this expression, the radial function~(\ref{horizonsolution}) reads
\begin{equation}\label{R}\
R(y) = C_1 e^{-\frac{i\omega}{2r_h^2} \ln y}\frac{\Gamma(c)\Gamma(c-a-b)}{\Gamma(c-a)\Gamma(c-b)} F_1+C_1 e^{-\frac{i\omega}{2r_h^2}  \ln
y}(1-y)^{c-a-b}\frac{\Gamma(c)\Gamma(a+b-c)}{\Gamma(a)\Gamma(b)}F_2~,
\end{equation}
and at infinity it can be written as
\begin{equation}\label{R2}\
R_{asymp.}(y) = C_1 \frac{\Gamma(c)\Gamma(c-a-b)}{\Gamma(c-a)\Gamma(c-b)}+C_1 (1-y)^{3/2}\frac{\Gamma(c)\Gamma(a+b-c)}{\Gamma(a)\Gamma(b)}= C_1 \frac{\Gamma(c)\Gamma(c-a-b)}{\Gamma(c-a)\Gamma(c-b)}~.
\end{equation}
So,
the field at infinity vanishes
if $c-a=-n$ or $c-b=-n$ for $n=0,1,2,...$.
Therefore, the QNFs are given by
\begin{equation}\label{w1}
\omega = -\frac{i \left((n+1) (2n+3) q^2 +\kappa^2\right)}{4n+5}~.
\end{equation}
Because all the QNFs have a purely negative imaginary part, we conclude that the massless scalar field perturbations are stable in the $z=2$ black hole background and that the system is always overdamped. Note that the QNFs for massless scalar fields do not depend on the spacetime dimension as we anticipated when the effective potential was analyzed.

\subsubsection{Scalar field stability for generic $z$ with Dirichlet boundary condition}

We observe from \eqref{w1} that the quasinormal frecuencies are purely imaginary and negative; however,  this result can be obtained by a more general argument which follows from \cite{Horowitz:1999jd}, adapted to Lifshitz geometries with a hyperscaling violation with generic values of $z$. So, by using infalling Eddington-Finkelstein coordinates $v=t+r^*$, and the ansatz
\begin{equation}
\varphi=e^{-i\omega v}e^{i\vec{\kappa} \cdot \vec{x}} \frac{\psi(r)}{r^s}\,,
\end{equation}
the Klein-Gordon equation yields
\begin{equation}\label{KleinFink}
\frac{d}{dr}(r^{1+z}f(r)\psi'(r))-2i\omega\psi'(r)-\frac{V(r)}{f(r)r^{z+1}}\psi(r)=0\,.
\end{equation}
Then, multiplying Eq. (\ref{KleinFink}) by $\psi^{\ast}$ and performing integrations by parts, and using the Dirichlet boundary condition for the scalar field at spatial infinity, one can obtain
\begin{equation}\label{relacion}
\int _{r_{h}}^{\infty}dr \left( r^{1+z}f(r) \left|  \frac{d\psi}{dr}\right|^2+\frac{V(r)}{f(r)r^{z+1}} \left| \psi \right| ^2 \right)=-\frac{\left|\omega \right|^2 \left| \psi (r=r_{h})\right| ^2}{\omega_{I}}\,,
\end{equation}
where $\omega_{I}$ is the imaginary part of $\omega$. In general, the QNFs are complex, where the real
part represents the oscillation frequency and the imaginary part describes the rate at which this oscillation is damped, with the stability of the scalar field being guaranteed if the imaginary part is negative.
Notice that the effective potential (\ref{pot}) is positive for $s>0$, which is guaranteed if $z>1$, then the left hand side of \eqref{relacion} is strictly positive, which demands that $\omega_{I}<0$ , and finally we conclude that the stability of the scalar field under perturbations respecting Dirichlet  boundary conditions is obeyed.

\subsubsection{Numerical analysis}
The Klein-Gordon equation (\ref{radial}) under
the following change of variable $y=1-r_{h}/r$ becomes
\begin{eqnarray}
\nonumber && \frac{d^2R}{dy^2}+\left(\frac{\eta-\tilde{\theta}}{1-y}+\frac{f^{\prime}(y)}{f(y)}\right) \frac{dR}{dy}\\
&&+\left( \frac{\omega^2(1-y)^{2z-2}}{r_h^{2z}f(y)^2}-\frac{\kappa ^2}{r_h^2 f(y)}-\frac{m^2 r_{h}^{\tilde{\theta}}}{(1-y)^{\tilde{\theta}+2} f(y)}\right)R=0\,.
\label{numericalmethod}
\end{eqnarray}
In this equation $f(y)$ refers to the function $f(r)$ evaluated at $r=\frac{r_{h}}{1-y}$; that is,
\begin{equation}
f(y)=1-\frac{q^{2}(1-y)^{2(z-1)}}{2(z-1)r_{h}^{2(z-1)}}\,,
\end{equation}
and $f'(y)=\frac{df(y)}{dy}$.
Now, we study the behavior of the radial equation (\ref{numericalmethod}) on the horizon and at the spatial infinite in order to consider the boundary conditions in both limits to then apply the improved AIM. 

At the event horizon, the effective potential $V(r)$ tends to zero; thus, the solution to the Schr\"{o}dinger equation is given by
\begin{equation}
F(r^{\ast})=C_{1}e^{-i\omega r^{\ast}}+C_{2}e^{i\omega r^{\ast}}\,,
\end{equation}
and imposing as a boundary condition that only ingoing waves exist on the event horizon, we must set $C_{2}=0$. Therefore, the solution near the horizon is given by
\begin{equation}\label{horizon}
F(r^{\ast})=C_{1}e^{-i\omega r^{\ast}}\,.
\end{equation}
On the other hand, the behavior of the effective potential at the spatial infinity ($r^* \rightarrow 0$) depends strongly on $\tilde{\theta}$ and $m$. We distinguish three cases:

\begin{itemize}

\item{$\tilde{\theta}<0$ or $m=0$ (with arbitrary $\tilde{\theta}$)}

In these cases the effective potential asymptotically tends to:
\begin{equation} \label{pott}
V(r^{\ast})=\frac{\delta}{z^2r^{\ast 2}}\,,
\end{equation}
where $\delta=s(s+z)=
(3z-1)(z-1)/4 >0$; therefore, the effective potential asymptotically tends to $+\infty$ ($z>1$).

\item{$\tilde{\theta} > 0$ and $m \neq 0$}

In this case the effective potential asymptotically tends to:
\begin{equation}
V(r^{\ast})=m^2 \left( -\frac{1}{z r^*} \right)^{2+\tilde{\theta}/z} \,.
\end{equation}
Notice that the effective potential asymptotically tends to $+\infty$, due to $2+\tilde{\theta}/z>0$.

\item{$\tilde{\theta} =0$ and $m \neq 0$}

In this case the effective potential is given by \eqref{pott}, but with $\delta=s(s+z)+m^2$.

\end{itemize}

In the numerical study we will perform below, we will consider only the first case (either $\tilde{\theta}<0$ or $m=0$), as the Schr\"{o}dinger equation at the spatial infinity can be solved for generic values of $\tilde{\theta}$ and $z$. In the second case, we were unable to solve the Schr\"{o}dinger equation at the spatial infinity 
 for generic values of $\tilde{\theta}$ and $z$.

Therefore, at spatial infinity, the solution to the Schr\"{o}dinger equation for $\tilde{\theta}<0$ or $m=0$ (with arbitrary $\tilde{\theta}$) is
\begin{equation}
F(r^{\ast})=D_{1}r^{ \ast \frac{1}{2}(1-\sqrt{1+\frac{4\delta}{z^2}})}+D_{2}r^{\ast \frac{1}{2}(1+\sqrt{1+\frac{4\delta}{z^2}})}. \\
\end{equation}
So, imposing the Dirichlet boundary condition, that is, having a null scalar field at spatial infinity, we must set $D_{1}=0$. Therefore, the solution becomes
\begin{equation}\label{infinity}
F(r^{\ast})=D_{2}r^{\ast \frac{1}{2}(1+\sqrt{1+\frac{4\delta}{z^2}})}\,.
\end{equation}

\begin{figure}[h]
\begin{center}
\includegraphics[width=0.9\textwidth]{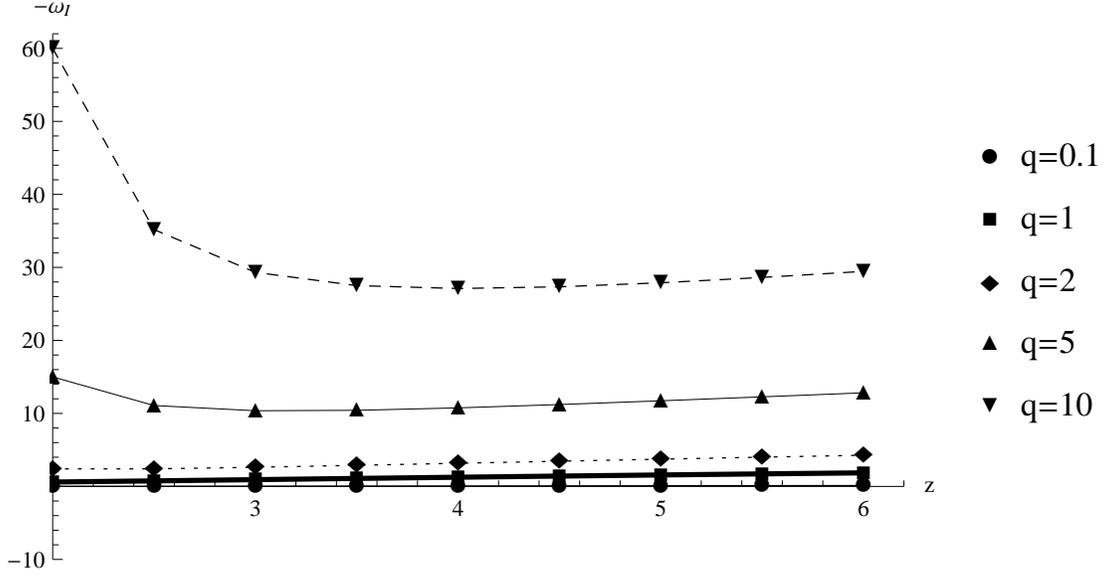}
\end{center}
\caption{The behavior of $-\omega_{I}$ of the fundamental quasinormal frequencies with $m=0$, $\kappa=0$ and different values of $q$ as a function of $z$.} \label{Fig}
\end{figure}

Thus, taking into account these behaviors, in terms of the variable $y$, where the event horizon is located at $y=0$ and the spatial infinity at $y=1$, we define
\begin{equation} \label{chieq}
R\left( y\right) = y^{-\frac{i\omega}{r_{h}^{z}f'(0)}} (1-y)^{\frac{1}{2}z(1+\sqrt{1+\frac{4\delta}{z^2}})+s} \chi (y)\,
\end{equation}
as ansatz. Then, by substituting these fields in Eq. (\ref{numericalmethod}) we obtain the homogeneous
linear second-order differential equation for the function $\chi (y)$
\begin{equation}
\chi ^{\prime \prime }=\lambda _{0}(y)\chi ^{\prime }+s_{0}(y)\chi \,,
\label{de}
\end{equation}%
where
\begin{eqnarray}
\notag \lambda _{0}(y)&=&\frac{1}{r_{h}^z f'(0)y(1-y)f(y)} (( r_{h}^z f'(0) y (3+2s+\tilde{\theta}-D(1+\tilde{\theta}/2)+z \sqrt{1+4 \delta/z^2}) \\
&&+ 2i (1-y) \omega )f(y)-r_{h}^z f'(0) y(1-y) f'(y)) \,, \\
\notag s_{0}(y)&=& -\frac{1}{2r_{h}^{2(z+1)}f'(0)^2y^2(1-y)^{2+\tilde{\theta}}f(y)^2}(2r_{h}^2f'(0)^2 \omega^2 y^2(1-y)^{2z+\tilde{\theta}}-r_{h}^2(1-y)^{\tilde{\theta}} \\
\notag &&( r_{h}^{2z}f'(0)^2y^2 (-2s^2-2\delta+(D-2)z(1+\tilde{\theta}/2)(1+\sqrt{1+4\delta/z^2})+2s((D-2)(1+\tilde{\theta}/2)\\
\notag &&-z\sqrt{1+4\delta/z^2}))-2ir_{h}^zf'(0)(1-y)
(1+2sy-(D-2)(1+\tilde{\theta}/2)y+z \sqrt{1+4\delta/z^2}y)\omega \\
\notag && +2(1-y)^2\omega^2)f(y)^2-r_{h}^zf'(0)yf(y)(2r_{h}^zf'(0)y(r_{h}^{\tilde{\theta}+2}m^2+\kappa^2 (1-y)^{2+\tilde{\theta}})+r_{h}^2 (1-y)^{1+\tilde{\theta}}\\
&&(r_{h}^zf'(0)(z(1+\sqrt{1+4\delta/z^2})+2s)y +2i(1-y)\omega)f'(y)))\,.
\end{eqnarray}
We solve this equation numerically (see Appendix A),
and we choose different values for the parameters. In Table \ref{QNM1}  we show fundamental QNFs  for massless scalar fields with different values of the dynamical exponent $z$ and different values of $q$. The QNFs are valid for the general spacetime dimension $D$ as the effective potential turns out to be independent of $D$ when $m=0$ and $\tilde{\theta} \neq 0$; however, for a fixed value of $z$ different values of $D$ implies different values of $\tilde{\theta}$ according to $\eqref{hyp}$. We have therefore also incorporated the values of the hyperscaling exponent in the second column of Table \ref{QNM1} for $D=4$. We can observe that $|\omega_I|$ increases when the charge $q$ increases. We also observe that for low values of $q$ ($q=0.1,1,2$) $|\omega_I|$ increases when the dynamical exponent increases, whereas for high values of $q$, for instance $q=10$, we can see that $|\omega_I|$ initially decreases, in the range $2\leq z \leq 4$,
and then increases when the dynamical exponent increases, in the range $4.5\leq z\leq 6.0$. For $q=5$, $|\omega_I|$ behaves in a similar manner.
This is also shown more explicitly in Fig. \ref{Fig}. On the other hand, by comparing the QNFs of the first row of Table \ref{QNM1} with the QNFs obtained analytically in the previous section, we conclude that the relative error is less than $\approx 5 \cdot 10^{-6}$ for $q=2$, and even less for the other values of $q$. Therefore, the improved AIM method is in good agreement with the $z=2$ analytical case.

Next, in Table \ref{QNM2}  we consider massless scalar fields with $\kappa=0$, $\tilde{\theta}=0$ (null hyperscaling factor), and different values of $z$ and $q$.  We observe that $|\omega_I|$ increases when the charge $q$ increases. We also observe that for low values of $q$, $|\omega_I|$ increases when $z$ increases, whereas for high values of $q$, $|\omega_I|$ initially decreases and then increases when the $z$ increases. This behavior is similar to that shown in Table \ref{QNM1}. In Table \ref{QNM3} we show some QNFs for the massless scalar field, $\kappa=0$, $q=1$, and different values of $z$, where we can observe a similar behavior for $|\omega_I|$. Finally, in Table \ref{QNM4}  we show fundamental QNFs for $\kappa=0$, $q=1$, $z=2$, and different values of $m$ and $D$. Here, we observe that $|\omega_I|$ increases when the scalar field mass increases, and for $m \neq 0$ $|\omega_I|$ increases with the spacetime dimension. Furthermore,  we observe that in all the cases analyzed the QNFs have an imaginary part that is negative and always overdamped, which ensures the stability  of the scalar perturbations in the background of a non-Abelian charged Lifshitz black hole with hyperscaling violation.
\begin{table}[ht]
\caption{Fundamental quasinormal frequencies for $m=0$, $\kappa=0$, and different values of $q$ and $z$. }
\label{QNM1}\centering
\begin{tabular}{ | c | c | c | c | c | c | c |}
\hline
$z$ & $\tilde{\theta} (D=4)$ & $q=0.1$ & $q=1$ & $q=2$ & $q=5$ & $q=10$\\ \hline
$2.0$ & $-1.0$ & $-0.00600 i$ &  $-0.60000 i$ & $-2.40001 i$ & $-15.00000 i$ & $-60.00010 i$  \\
$2.5$ & $-0.5$ & $-0.01633 i$ &  $-0.75795 i$ & $-2.40634 i$ & $-11.08130 i$ & $-35.18090 i$ \\
$3.0$ & $0$ & $-0.02932 i$ &  $-0.92725 i$ & $-2.62266 i$ & $-10.36700 i$ & $-29.32230 i$ \\
$3.5$ & $0.5$ & $-0.04360 i$ &  $-1.09505 i$ & $-2.88985 i$ & $-10.42300 i$ & $-27.50630 i$  \\
$4.0$ & $1.0$ & $-0.05840 i$ &  $-1.25824 i$ & $-3.17057 i$ & $-10.75780 i$ & $-27.10800 i$  \\
$4.5$ & $1.5$ & $-0.07335 i$ &  $-1.41608 i$ & $-3.45243 i$ & $-11.21400 i$ & $-27.34010 i$   \\
$5.0$ & $2.0$ & $-0.08820 i$ &  $-1.56852 i$ & $-3.73059 i$ & $-11.72740 i$ & $-27.89260 i$   \\
$5.5$ & $2.5$ & $-0.10286 i$ &  $-1.71581 i$ & $-4.00307 i$ & $-12.26770 i$ & $-28.62140 i$ \\
$6.0$ & $3.0$ & $-0.11725 i$ &  $-1.85826 i$ & $-4.26915 i$ & $-12.81950 i$ & $-29.45140 i$ \\ \hline
\end{tabular}%
\end{table}

\begin{table}[ht]
\caption{Fundamental quasinormal frequencies for $m=0$, $\kappa=0$, $\tilde{\theta}=0$, and different values of $z$ and $q$. }
\label{QNM2}\centering
\begin{tabular}{ | c | c | c | c | c | c | c | }
\hline
$z$ & $D=z+1$ & $q=0.1$ & $q=1$ & $q=2$ & $q=5$ & $q=10$\\ \hline
$3$ & $4$ & $-0.02932 i$ &  $-0.92725 i$ & $-2.68266 i$ & $-10.36700 i$ & $-29.32230 i$  \\
$4$ & $5$ &  $ -0.05840 i$ &  $-1.25824 i $ & $-3.17057 i$ & $-10.75780 i$ & $-27.10800 i$ \\
$5$ &  $6$ & $-0.08820 i$ &  $-1.56852 i$ & $-3.73059 i$ & $-11.72740 i$ & $-27.89260 i$ \\
$6$ & $7$ &  $-0.11725 i$ &  $-1.85826 i$ & $-4.26915 i$ & $-12.81950 i$ & $-29.45140 i$  \\
$7$ & $8$ &  $-0.14511 i$ &  $-2.12998 i$ & $-4.78164 i$ & $-13.92650 i$ & $-31.26380 i$  \\ \hline
\end{tabular}%
\end{table}

\begin{table}[ht]
\caption{Lowest quasinormal frequencies for $m=0$, $\kappa=0$, $q=1$, $D=4$ and different values of $z$. }
\label{QNM3}\centering
\begin{tabular}{ | c | c | c | c | c | c |}
\hline
$z$ & $\tilde{\theta}(D=4)$ & $n=0$ & $n=1$ & $n=2$ & $n=3$   \\ \hline
$2$ & $-1$ & $-0.60000 i$ & $-1.11111 i$ & $-1.61539 i$ & $-2.11765 i$  \\
$3$ & $0$ & $-0.92725 i$ & $-1.64194 i$ & $-2.35195 i$ & $-3.06051 i$  \\
$4$ & $1$ & $-1.25824 i$ & $-2.15022 i$ & $-3.06389 i$ & $-3.97281 i$  \\ \hline
\end{tabular}%
\end{table}

\begin{table}[ht]
\caption{Fundamental quasinormal frequencies for $\kappa=0$, $q=1$, $z=2$, and different values of $m$ and $D$. }
\label{QNM4}\centering
\begin{tabular}{ | c | c | c | c |}
\hline
$m$ & $D=4$ & $D=5$ & $D=6$ \\
$$ & $\tilde{\theta}=-1$ & $\tilde{\theta}=-4/3$ & $\tilde{\theta}=-3/2$ \\ \hline
$0$ & $-0.60000 i$ & $-0.60000 i$ & $-0.60000 i$ \\
$1$ & $-0.72345 i$ & $-0.74414 i$ & $-0.75601 i$ \\
$2$ & $-1.01120 i$ & $-1.09099 i$ & $-1.14103 i$ \\
$3$ & $-1.35329 i$ & $-1.50305 i$ & $-1.59933 i$ \\
$4$ & $-1.71014 i$ & $-1.92785 i$ & $-2.06783 i$ \\
$5$ & $-2.07227 i$ & $-2.35645 i$ & $-2.53907 i$ \\
$6$ & $-2.43685 i$ & $-2.78670 i$ & $-3.01148 i$ \\
$7$ & $-2.80276 i$ & $-3.21783 i$ & $-3.48450 i$ \\
$8$ & $-3.16946 i$ & $-3.64948 i$ & $-3.95787 i$ \\
$9$ & $-3.53668 i$ & $-4.08145 i$ & $-4.43146 i$ \\ \hline
\end{tabular}%
\end{table}

\section{Final remarks}
\label{conclusion}
In this work we studied the QNFs of scalar field perturbations of non-Abelian hyperscaling violating Lifshitz black holes.
First, we obtained analytically the QNFs of massless scalar fields for Lifshitz black holes with dynamical exponent $z=2$ and  hyperscaling violating factor $\tilde{\theta}>-2$. Then,  we considered two cases: $-2<\tilde{\theta}<0$ and $m=0$ (with arbitrary $\tilde{\theta}$), and we obtained the QNFs numerically.
We also studied the stability of these massive and massless scalar field perturbations on the black holes under consideration through the QNFs. Our results show that the QNFs are purely imaginary and negative; therefore, the perturbation is always overdamped. It is worth mentioning that for other geometries with a hyperscaling violating exponent \cite{Gonzalez:2015gla}, there is a limit on the dynamical exponent above which the perturbations are always overdamped for a given dimension, and the hyperscaling violating exponent shifts this limit. In fact, in \cite{Gonzalez:2015gla} it was found that the QNMs are always overdamped for $D \leq z+2+\theta$ and are non-overdamped otherwise in all the cases analyzed numerically. As we mentioned, the exponent $\theta$ is related to $\tilde{\theta}$ by $\theta=-(D-2) \tilde{\theta} /2$, so, in terms of $\tilde{\theta}$ the above inequality reads $D \leq z+2-(D-2) \tilde{\theta} /2$. Furthermore, taking into account that $\tilde{\theta}$ is given by Eq. \eqref{hyp}, we can see that our results are in agreement with this law, because the QNMs are always overdamped. Also, it is worth mentioning that the QNFs for massless scalar fields do not depend on the spacetime dimension because the effective potential is independent of $D$ in that case, which can be seen in the analytical result given in Eq. \eqref{w1} as well as in the numerical results given in the Tables. In addition, according to the gauge/gravity duality, the relaxation time $\tau$ for a thermal state to reach thermal equilibrium in the boundary conformal field theory is $\tau=1/ |\omega_I |$, where $\omega_I$ is the imaginary part of the fundamental QNF. So, for stable configurations it is  possible to reach thermal equilibrium and, as can be deduced from Tables \ref{QNM1}-\ref{QNM4}, in general, we found that when the dynamical exponent increases, and thus the hyperscaling violating exponent increases for a given dimension according to Eq. \eqref{hyp}, the relaxation time of the dual thermal states decreases for low values of the charge $q$; however, for high values of $q$, when the dynamical exponent increases the relaxation time of the dual thermal states initially increases and then decreases.
Also, when the charge $q$ and/or the mass of the scalar field increases the relaxation time decreases. This behavior is contrary to what was obtained in \cite{Gonzalez:2015gla,Sybesma:2015oha}, where the relaxation time always increased.

\newpage

\acknowledgments

We would like to thank the anonymous referees for the very useful comments which help us improve the quality of our paper. This work was partially funded by the Comisi\'{o}n
Nacional de Ciencias y Tecnolog\'{i}a through FONDECYT Grant 11140674 (PAG) and by the Direcci\'{o}n de Investigaci\'{o}n y Desarrollo de la Universidad de La Serena (Y.V.). P. A. G. acknowledges the hospitality of the Universidad de La Serena where part of this work was undertaken. R.B. acknowledges the hospitality of the Universidad Diego Portales.

\appendix
\section{Improved AIM}
In this appendix we give a brief review of the improved AIM, which is used to solve
homogeneous
linear second-order differential equations subject to boundary conditions. First, it is necessary to implement the boundary conditions. For this purpose the dependent variable must be redefined in terms of a new function, say $\chi$, that satisfies the boundary conditions appropriate to the eigenvalue problem under consideration. In the study of quasinormal modes of the black holes one solves the radial equation on the horizon and at spatial infinity, and imposes the boundary condition that on the horizon only ingoing waves exist there and at spatial infinity the appropriate boundary condition depends on the asymptotic behavior of the spacetime (in our case we imposed the scalar field to be null at spatial infinity due to the effective potential diverges there; therefore, the new radial function was defined in Eq. (\ref{chieq})). Thus, in order to implement the improved AIM the differential equation must be written in the form
\begin{equation}
\chi ^{\prime \prime }=\lambda _{0}(y)\chi ^{\prime }+s_{0}(y)\chi \,.
\label{de}
\end{equation}%
Then,
one must differentiate Eq. (\ref{de}) $n$ times with respect to $y$,
which yields the following equation:
\begin{equation}
\chi ^{n+2}=\lambda _{n}(y)\chi ^{\prime }+s_{n}(y)\chi~,  \label{de1}
\end{equation}%
where
\begin{equation}
\lambda _{n}(y)=\lambda _{n-1}^{\prime }(y)+s_{n-1}(y)+\lambda
_{0}(y)\lambda _{n-1}(y)~,  \label{Ln}
\end{equation}%
\begin{equation}
s_{n}(y)=s_{n-1}^{\prime }(y)+s_{0}(y)\lambda _{n-1}(y)\,.  \label{Snn}
\end{equation}%
Then, expanding the $\lambda _{n}$ and $s_{n}$ in a Taylor series around
some point $\eta $, at which the improved AIM is performed, yields
\begin{equation}
\lambda _{n}(\eta )=\sum_{i=0}^{\infty }c_{n}^{i}(y-\eta )^{i}\,,
\end{equation}%
\begin{equation}
s_{n}(\eta )=\sum_{i=0}^{\infty }d_{n}^{i}(y-\eta )^{i}\,,
\end{equation}%
where the $c_{n}^{i}$ and $d_{n}^{i}$ are the $i^{th}$ Taylor coefficients
of $\lambda _{n}(\eta )$ and $s_{n}(\eta )$, respectively, and by substituting
the above expansions in Eqs. (\ref{Ln}) and (\ref{Snn}) the following
set of recursion relations for the coefficients is obtained:
\begin{equation}
c_{n}^{i}=(i+1)c_{n-1}^{i+1}+d_{n-1}^{i}+%
\sum_{k=0}^{i}c_{0}^{k}c_{n-1}^{i-k}~,
\end{equation}%
\begin{equation}
d_{n}^{i}=(i+1)d_{n-1}^{i+1}+\sum_{k=0}^{i}d_{0}^{k}c_{n-1}^{i-k}\,.
\end{equation}%
Thus, the authors of the improved AIM have avoided the
derivatives that contain the AIM in \cite{Cho:2009cj, Cho:2011sf}, and
the quantization condition, which is equivalent to imposing a termination
on the number of iterations, is given by
\begin{equation}
d_{n}^{0}c_{n-1}^{0}-d_{n-1}^{0}c_{n}^{0}=0~.
\end{equation}

\end{document}